# Interfacial free energy and Tolman length of curved liquid-solid interfaces from equilibrium studies

P. Montero de Hijes,[1] Jorge. R. Espinosa,[2, 3] Valentino Bianco,[1] Eduardo Sanz,[1] and Carlos Vega[1]

[1] *Departamento de Química Física, Facultad de Ciencias Químicas, Universidad Complutense de Madrid, 28040 Madrid, Spain*

[2] *Maxwell Centre, Cavendish Laboratory, Department of Physics, University of Cambridge, Cambridge CB3 0H3, United Kingdom*

[3] *Emmanuel College, Cambridge, CB2 3AP, United Kingdom*

In this work, we study by means of simulations of hard spheres the equilibrium between a spherical solid cluster and the fluid. In the NVT ensemble we observe stable/metastable clusters of the solid phase in equilibrium with the fluid, representing configurations that are global/local minima of the Helmholtz free energy. Then, we run NpT simulations of the equilibrated system at the average pressure of the NVT run and observe that the clusters are critical because they grow/shrink with a probability of 1/2. Therefore, a crystal cluster equilibrated in the NVT ensemble corresponds to a Gibbs free energy maximum where the nucleus is in unstable equilibrium with the surrounding fluid, in accordance with what has been recently shown for vapor bubbles in equilibrium with the liquid. Then, within the Seeding framework, we use Classical Nucleation Theory to obtain both the interfacial free energy $\gamma$ and the nucleation rate. The latter is in very good agreement with independent estimates using techniques that do not rely on Classical Nucleation Theory when the mislabeling criterion is used to identify the molecules of the solid cluster. We therefore argue that the radius obtained from the mislabeling criterion provides a good approximation for the radius of tension, $R_s$. We obtain an estimate of the Tolman length by extrapolating the difference between $R_e$ (the Gibbs dividing surface) and $R_s$ to infinite radius. We show that such definition of the Tolman length coincides with that obtained by fitting $\gamma$ versus $1/R_s$ to a straight line as recently applied to hard spheres [Montero de Hijes et al., J. Chem. Phys. 151, 155401, 2019].

## I. INTRODUCTION

The thermodynamics of systems having two phases with a curved interface is a fascinating topic that has been largely discussed by the scientific communities in the last decades[1–18]. A system with a fixed number of particles ($N$), volume ($V$), and temperature ($T$) can exhibit a stable/metastable spherical interface between the solid and liquid phase corresponding to a global/local minimum of the Helmholtz free energy ($F$)[2,5,15,19–30]. Thermodynamic properties of metastable states can be studied as long as there is a free energy barrier separating them from the equilibrium one, and the relaxation time of the system is shorter than the time required to overcome the free energy barrier[31–33]. According to the thermodynamic description of Rowlinson and Widom[34] for planar interfaces at equilibrium, the value of the interfacial free energy $\gamma$ is unique while for curved interfaces depends on the choice of the dividing surface between the two phases[7,34]. There are two reasonable choices: the Gibbs dividing surface with radius $R_e$ and surface free energy $\gamma_e$ (where the excess number of particles is zero), and the surface of tension with radius $R_s$ and surface free energy $\gamma_s$ satisfying the Laplace equation, which for spherical interfaces reads $\Delta p = 2\gamma_s/R_s$, being $\Delta p$ the pressure difference across the interface[35].

Whenever thermodynamics enters in action, one can also use Statistical Mechanics to get a microscopic insight. In fact, Kirkwood and Buff have shown that for a planar interface between fluid phases, it is possible to evaluate $\gamma$ (which is unique) from a mechanical route by computing the pressure tensor[36]. This approach has been adopted in several simulation works, following the pioneering study of Chapela et al.[37]. However, there are cases where there is no rigorous mechanical route to $\gamma$, including the planar fluid-solid[38,39], curved fluid-fluid, and curved solid-fluid[7,10,34,40] interfaces. The only way to calculate $\gamma$ in these cases requires the evaluation of the total Helmholtz free energy of the system $F$. Not surprisingly, the lack of a mechanical route to $\gamma$ results in quite infrequent experimental approaches to measure $\gamma$ for planar fluid-solid interfaces (ice-water interface being an excellent example of the situation[41,42]), if not entirely absent or dubiously rigorous as in the case of curved interfaces.

After this frustrating situation, several routes to $\gamma$ for curved interfaces have been proposed. The first route consists in assuming that the value of $\gamma$ for the curved interface is that of the planar interface. This is denoted as the capillarity approximation. The approach is simple, but there is no fundamental reason to believe that the value of $\gamma$ does not depend on the curvature of the interface[13,43]. Indeed, a series of studies on nucleation phenomena have provided indirect evidences that the capillarity approximation fails, as $\gamma$ changes with the curvature of the spherical phase[31,44–49].

The second route is theoretical and was initiated by Tolman. In 1949, Tolman wrote an influential paper entitled "The effect of droplet size on surface tension"[50]. He assumed that the difference $\delta$ between the curvature radii $R_e$ and $R_s$ of the two dividing surfaces is constant (i.e. it does not change with the radius of the spherical cluster) and that can be estimated by its value for infinitely large clusters (which defines the Tolman length $\delta_{Tolman}$). By performing certain approximations he showed that the value of $\gamma$ (along an isotherm) should change with the the inverse of $R_s$ with an expression where the Tolman length plays a key role. The paper provides molecular evidences that $\gamma$ changes with the radius of curvature of the cluster, shifting the discussion to the value of $\delta_{Tolman}$ characterizing the difference between $R_e$ and $R_s$ for infinitely large clusters. Determining $R_e$ is quite simple and only requires to know the bulk densities of the two phases and the total volume (the detailed density profile is not needed). However, the absence of a rigorous mechanical method to obtain $R_s$ (in spite of a series of attempts[7,10,34,40]) implies that it can only be determined rigorously through the cumbersome task of determining the Helmholtz free energy of the system $F$. Since rigorous calculations of $F$ for systems with curved interfaces are commonly missing, $R_s$ is not determined rigorously. This has generated an intense debate on the magnitude and sign of the Tolman length for a number of systems for more than 70 years[1–18,51].

The third route to obtain $\gamma$ for curved interfaces was initiated by Turnbull and co-workers[52,53]. It uses nucleation studies to estimate values of $\gamma$ for curved interfaces being this an indirect route. The idea of Turnbull was to fit the experimental values of the homogeneous nucleation rate $J$ (i.e. number of critical clusters per unit of time and volume) using classical nucleation theory (CNT) which can be regarded as a combination of the formalism of Volmer-Weber[54]–Becker-Doring[55] and the Gibbsian formalism applied to the thermodynamics of curved interfaces[56]. This interesting approach takes advantage of experimental results for $J$ to obtain, after a theoretical treatment, values of $\gamma$ for curved interfaces[57]. This route has undergone a new revival from simulation studies in the last decade as now it is possible to estimate $J$ for a potential model using computer simulations. Techniques like umbrella sampling[44,45] (US), forward flux sampling[46] (FFS), or transition path sampling[58] (TPS) can be used to determine $J$. These techniques are rigorous but rather expensive from a computational point of view. For this reason, in the last ten years, a new technique has been proposed aimed to determine $J$ denoted as Seeding[59–65], where a solid cluster (equilibrated at a certain value of $T$ and $p$) is inserted into an equilibrated liquid (at the same conditions $T, p$) to determine whether is critical or not. According to its time evolution in the $NpT$ ensemble: it is critical if the probability to freeze and to melt are equal while there is no other possible result than these two options. The methodology does not allow to estimate $J$ by itself. However, following the CNT with a "judicious" choice of the order parameter[66] used to determine the size of the critical cluster, one can reasonably estimate free energy barriers and nucleation rates getting values comparable with the ones obtained with rigorous techniques[62,67,68]. The

Seeding method also provides (through the CNT formalism) values of $\gamma$ for the curved interface. The Seeding scheme resembles Turnbull's approach in the sense that it connects nucleation studies and CNT (Turnbull's approach going from $J$ to $\gamma$ using experimental results of $J$, and Seeding going from $\gamma$ to $J$ using simulation results). Interestingly, we have shown recently that the values of $\gamma$ from our Seeding studies of nucleation, can be described by a "Tolman-like" expression for a number of systems including HS[49]. An interesting question (that we intend to address in this work) is whether this Tolman-like expression is also able to describe results for curved interfaces at equilibrium.

Let us now discuss the fourth route to $\gamma$ for a curved interface. It simply requires to study a system that is at equilibrium and that presents a curved interface. This route has been developed for simulations studies by Binder and coworkers[2,5,19–27], showing that in a system at constant $N$, $V$, and $T$ it is possible to have fluid-solid configurations with curved interfaces in equilibrium, corresponding to a minimum of $F$. Depending on the $(N, V, T)$ conditions, the minimum of $F$ corresponds to i) a sphere of the solid phase within the fluid; ii) an infinite cylinder of the solid phase (percolating through the periodical boundary conditions) in contact with the fluid; iii) a slab of the solid phase in contact with the fluid; iv) all the previous cases switching the roles of the fluid and solid phases. At some point, the minimum of $F$ may correspond to an homogeneous fluid or homogeneous solid phase.

By focusing on the vapor-liquid interfaces, Binder and coworkers[2,5,21,23,27] evaluated $F$, and determined (for each considered system) the value of R for which $\gamma$ was minimum (thus obtaining $\gamma_s$ and $R_s$). They observed that the capillarity approximation does not work (i.e. $\gamma_s$ changed with $R_s$ ) and also that the difference between $R_e$ and $R_s$ was not constant either. So far, these studies focused on the liquid-vapor interface are the only rigorous route so far to $\gamma$ for a curved interface and can be considered as a tour de force. After all, determining $F$ in computer simulations is possible but terribly expensive. For the fluid-solid interface, there have been simulation studies showing that a spherical cluster may be stable (or metastable)[14,15,22,24,25] although a rigorous determination of $F$, to the best of our knowledge, is still missing.

Let us now present the main goals of this work. In this work, we address the issue of the variation of $\gamma$ with $R$ for a curved fluid-solid interface. Aiming to provide a rationale for some fundamental aspects of this intriguing problem, we will focus on a simple and pedagogical system: hard spheres (HS). We will show that for HS it is possible to obtain stable spherical clusters of the solid phase when the system is simulated in the canonical ensemble (NVT). Our findings are consistent with a recent study of Richard et al.[14]. We then show that the clusters equilibrated in the NVT ensemble are critical when the system is run at constant NpT, being $p$ the mean pressure of the NVT simulation. Knowing that the clusters are critical, we estimate the value of $\gamma_s$ for the clusters using the CNT approximations previously used in Seeding studies[49,62]. We get consistent values with our previous Seeding work in the NpT ensemble[62]. Since, as previously shown[62], the nucleation rate estimated using CNT for the simulated clusters is consistent with that obtained from independent techniques that do not rely on CNT, we identify the cluster radius obtained in the simulations with $R_s$. This identification enables two different routes to estimate the Tolman length. One is to extrapolate the difference between $R_e$ and $R_s$ (marking the distance between the equimolar dividing surface and the surface of tension) to infinitely large clusters. The other one is to linearly fit $\gamma_s$ vs $1/R_s$. We show in this paper that both definitions are consistent with each other as anticipated by Tolman.

## II. SIMULATION DETAILS

In this work we shall not study a true HS system, but rather a pseudo hard sphere system (PHS)[69]. The main reason is that the PHS potential is continuous. An advantage of having a continuous potential is that one can use highly efficient codes as GROMACS[70] (highly optimized for parallel calculations). A good choice is represented by the truncated and shifted Mie potential with power m=50 for the repulsion and n=49 for the attraction, called also pseudo hard sphere potential[69]:

$$u_{PHS}(r) = \begin{cases} 50 \left(\frac{50}{49}\right)^{49} \epsilon \left[\left(\frac{\sigma}{r}\right)^{50} - \left(\frac{\sigma}{r}\right)^{49}\right] + \epsilon & r < \left(\frac{50}{49}\right)\sigma \\ 0 & r \geq \left(\frac{50}{49}\right)\sigma \end{cases}$$

(1)

where $\sigma$ represents the hard sphere diameter and $\epsilon$ is the depth of the potential. $u_{PHS}$ reproduces almost exactly the properties of HS, like the equation of state, the diffusion coefficient, the glass transition, the phase diagram and, last but not least, the coexistence crystal-fluid interfacial free energy (which plays a major role in this work)[49,69,71–73]. The potential given by Eq.(1) is provided in Tabular form to GROMACS. We adopt the following parameters: $\sigma = 0.3405$ nm, $\epsilon/k = 119.87$ K ($k$ is the Boltzmann constant), and the particle mass $m = 6.69 \cdot 10^{-26}$ kg. These parameters are taken from the standard Lennard-Jones potential used to describe Ar. Simulations are performed at $T = 179.8$ K, since the properties reproduced by $u_{PHS}$ potential agree with the ones of HS at the reduced temperature $T^* = T/(\epsilon/k) = 1.5$. Integration time steps is fixed to 1.0 fs and the $T$ is kept constant by using the Nose-Hoover thermostat. In the following, we convert the real units of GROMACS to reduced units: $\sigma$ is the unit length; $\tau = \sigma\sqrt{m/(kT)}$ is the unit time (corresponding to 1.761ps); and $kT$ is the unit energy. According to this, the volume is expressed as $V^* = V/\sigma^3$, the density as $\rho^* = (N/V)\sigma^3$, the pressure as $p^* = p/(kT/\sigma^3)$, and the interfacial free energy as $\gamma^* = \gamma/(kT/\sigma^2)$. Hereafter, all the quantities writ-ten with a start as superscript will refer to quantities expressed in reduced HS units. In what follows, we shall denote the PHS model, simply as HS.

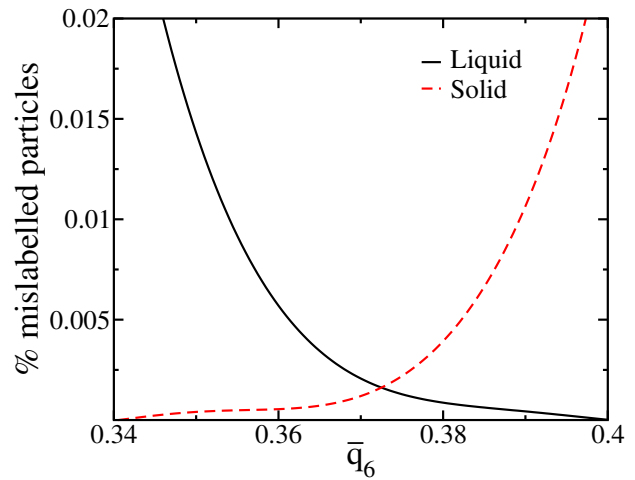

FIG. 1. Percentage of mislabeled particles in the bulk liquid and bulk solid for PHS at $p^* = 12.887$. The crossing point determines the threshold which in this work will be of $q_{\bar{6},t} = 0.372$. Molecules with $\bar{q}_6 > 0.372$ will be labeled as solid, whereas those with $\bar{q}_6 < 0.372$ will be labeled as liquid.

Each particle of the system is labeled as fluid or solid according to the Lechner-Dellago order parameter $\bar{q}_6$[74]. The threshold value of $q_{\bar{6},t}$ used to label each particle as liquid-like and solid-like was determined using the mislabeling criterium[61–63,68]. The mislabeling criterium states that the threshold value of the order parameter used to label particles as liquid or solid is obtained by simulating the bulk fluid and bulk solid phases and equating the small percentage of particles that are mislabeled as solid in the bulk fluid, to those that are mislabeled as liquid in the bulk solid. Particles at a distance of $1.33\sigma$ of a central one are considered neighbors. In Fig. 1, the mislabeling curves of HS in the fluid and solid phases are presented for the reduced pressure $p^* = p/(kT/\sigma^3) = 12.887$. From the curves, we adopt $q_{\bar{6},t} = 0.372$ as threshold value, checking that $q_{\bar{6},t}$ variations upon pressurization are negligible, within the pressure range explored in this work. Once each particle of the system is labeled as liquid or solid we shall evaluate the size of the largest solid cluster, being $N_{sol}$ the number of solid particles it contains.

For the adopted model, we have determined the value of the coexistence pressure[72] which is $p^* = 11.648$ (for true HS the value is of about 11.57, see Ref.[75]). Most of the simulations of this work lasted around 20ns. This timescale is (roughly) more than a thousand times the time required to diffuse a particle diameter (which for a pressure of $p^* = 12.5$ is of about 13.5ps, or $\sim 8\tau$ in reduced units of time).

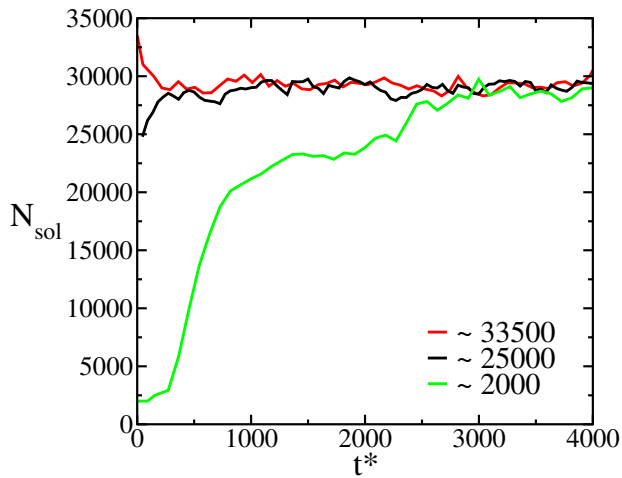

FIG. 2. Number of particles of the largest solid cluster in the system as function of time ($t^* = t/\tau$), for $N = 105875$ and $V^* = 108265.2$. Systems only differ on the initial size of the crystal seed.

| Label | $V^*$ | $N$ | $\langle N_{sol} \rangle$ | $\langle N_{sol} \rangle / N$ | $\langle p^* \rangle$ |
|-------|-------|-----|---------------------------|-------------------------------|-----------------------|
| I | 10686.4 | 10540 | 1925 | 0.183 | 13.209 |
| II | 20195.5 | 19779 | 2736 | 0.138 | 13.027 |
| III | 20195.5 | 19829 | 3718 | 0.188 | 12.887 |
| IV | 49599.9 | 48207 | 5604 | 0.116 | 12.739 |
| V | 49599.9 | 48357 | 8602 | 0.178 | 12.579 |
| VI | 108265.2 | 104675 | 10498 | 0.100 | 12.517 |
| VII | 66900.1 | 65383 | 15554 | 0.238 | 12.399 |
| VIII | 108265.2 | 105475 | 23558 | 0.223 | 12.306 |
| IX | 108265.2 | 105875 | 28879 | 0.273 | 12.258 |
| X | 887000.0 | 853712 | 129926 | 0.152 | 12.011 |

TABLE I. Thermodynamic variables of the closed finite systems simulated at constant temperature and volume. The pressure $\langle p^* \rangle$ is the average value obtained when the system reaches the equilibrium state. Only spherical clusters were considered. The average number of particles in the solid cluster is denoted as $\langle N_{sol} \rangle$.

## III. PHASE EQUILIBRIUM ABOVE COEXISTENCE: SOLID CLUSTERS STABILIZED IN LIQUIDS

We perform simulations at constant number of particles $N$, volume $V$, and temperature $T$. We seed a preformed spherical solid cluster of a certain size in the bulk liquid and let the system evolve towards the equilibrium. As shown in Fig. 2, different runs with the same values for $N$ (total number of particles in the system) and $V$ (volume of the simulation box), but with initial clusters differing in size, converge towards the same equilibrium state, where the solid cluster, in average, has the same size in all cases. When the initial cluster is rather small compared to the equilibrium one, it takes some time to reach the final size (green curve in Fig. 2). On the other hand, equilibration is much faster when the initial size is close to the equilibrium one. It should be pointed out that the size of the initial cluster cannot be chosen arbitrarily. Indeed, if the initial cluster is too small, it may melt; if it is too large then it may percolate through the simulation box forming a cylinder (we shall come to this point later). It should be considered also that some defects (of kinetic origin due to the fast growth) may arise when the cluster grows from a very small initial size. For this reason, it is advisable to use initial (and perfect) solid clusters close in size to the final equilibrium state. Once the equilibrium state is reached, there are thermal oscillations in the size of the solid cluster (Fig. 2) due to capillary waves fluctuating in the solid-fluid interface. Also, it should be noticed that it is convenient to consider the shape of the seed for technical reason albeit irrelevant in terms of stability. This is, if instead of a spherical cluster we inserted a cubic cluster, this would turn into a sphere-like cluster as soon as possible because the cubic is not even a local minimum.

We have repeated this procedure ten times, varying $N$ and $V$ ($T$ is constant and for hard spheres it just scales the velocities of the particles but does not affect configurational properties), obtaining in all cases equilibrium solid clusters of spherical average shape. The results are presented in Table I, where the average size of these solid equilibrium clusters (labeled as $\langle N_{sol} \rangle$) ranges from about 2000 up to 130000 particles. Notice that the size of the equilibrium cluster is uniquely determined by the values of $N$ and $V$, and corresponds to a minimum in the Helmholtz free energy $F$. Both the values of $\langle N_{sol} \rangle$ and $\langle N_{sol} \rangle / N$ obtained for a certain value of $N$ and $V$, are dictated by thermodynamics (i.e. the minimum in $F$) and cannot be changed at will. As can be seen in Table I, we found the ratio $\langle N_{sol} \rangle / N$ to be $\in [0.1 : 0.27]$. Although not stated explicitly in the books describing the thermodynamic treatment of curved interfaces, one has the impression that it is assumed that the volume of the fluid phase is many times larger than the volume of the solid phase. At least for HS, under periodical boundary conditions this is certainly not the case.

In Table I, we present also the values of the pressure obtained in the simulation runs during the period in which the solid cluster is stable. They fall in the range $p^* \in [12.011 : 13.209]$. Since the coexistence pressure is $p^* = 11.648$, our findings suggest that this equilibrium method applies only close to the coexistence.

By comparing the cases II and III in Table I, both sharing the same volume $V^* = 20195.5$, we see that by reducing $N$ the equilibrium solid cluster becomes smaller. The size of the equilibrium cluster is very sensitive to $N$. In fact, for clusters II and III, removing just 50 particles makes a change of about 1000 particles in $\langle N_{sol} \rangle$. Again, focusing on the isochoric cases VI, VIII and IX, the reduction of 1200 particles induces a change in $\langle N_{sol} \rangle$ of about 20000. As can be seen, decreasing the global density causes an increase of the volume of the phase with lower density (the fluid phase in this case) reducing the size of the solid cluster as given by $\langle N_{sol} \rangle$. Concerning the size of the equilibrium cluster we found that, while it is always possible to stabilize a cluster with a size as big as desired, the smallest equilibrium cluster we could obtain was composed by $\sim 2000$ particles. Below this threshold the solid clusters melted leading to the conclusion that for HS, within this method it is not possible to equilibrate solid clusters with much less than $\sim 2000$ particles.

In Fig. 3, we present the time evolution of $N_{sol}$ for the ten systems considered in Table I. Each panel contains trajectories with the same constant value of $V$ although different constant value of $N$. As can be seen in Fig. (3)a, during a short time the initial seed grows until it reaches a stable quasi-spherical size, maintained for a significant period of time ($\tau \in [500 : 2500]$), corresponding to about $\sim 250$ diffusion times. At larger times, $\tau > 2500$, the system undergoes a transition to a new conformation represented by a cylindrical solid.

In Fig. (4) we report two snapshots of the solid cluster before and after the transition. The fact that the spherical solid clusters is stable for certain time and that the change to the cylindrical shape occurs rapidly indicates that there is a free energy barrier separating the spherical cluster from the cylindrical one. In this case, the spherical cluster represents a local minimum of $F$ (a metastable configuration), while the cylinder represents a deepest (possibly global) minimum of $F$ (we never observed the transition from a cylinder to a sphere). The same transition was observed for cases III and IX, shown in Figs. (3)b and (3)e respectively.

## IV. CONNECTING EQUILIBRIUM AND NUCLEATION

We shall now perform an interesting exercise. We shall perform NpT simulations at the average pressure found in the NVT run (denoted as $\langle p^* \rangle$ in Table I). For the starting configuration, we shall randomly select one from the NVT run in which the cluster was stable. Then, we shall study in detail the time evolution of the solid cluster by launching up to 30 independent $NpT$ simulations (by changing the initial velocities). These results for systems III and VII of Table I are presented in Fig. 5, panels a) and b) respectively. The trajectories show that around half of the times the clusters melt and the other half, they grow until crystallizing the entire system. Hence, the clusters are critical at pressure $\langle p^* \rangle$. The fact that the nucleus equilibrated in the NVT ensemble is a critical cluster in the NpT ensemble has been recently shown for bubble cavitation in the Lennard Jones system[30]. In Fig. 6 we show the number of particles in the solid clusters versus $\langle p^* \rangle$ at equilibrium as obtained in this work in the NVT ensemble (red dots) alongside our previous results for the size of the critical clusters obtained in the NpT ensemble (black dots)[49,62]. As can be seen there is excellent agreement between both set of results further reinforcing the connection between equilibrium and nucleation.

When the cluster is critical, it must be in the top of a Gibbs free energy ($G$) curve. In other words, $G$ reaches a maximum when plotted as a function of the number of particles in the solid cluster while keeping $N$, $p$, and $T$ constant. However, the same system was in equilibrium when it was studied at constant N, V and T. Therefore, $F$ reaches a minimum when plotted as a function of the number of particles in the solid cluster while keeping $N$, $V$, and $T$ constant. That indicates that changing the number of particles in the solid cluster would increase the value of the Helmholtz free energy. This is sketched in Fig. 7.

The consequences of the results of Fig. 5 and Fig. 7 are important. On the one hand, if a thermodynamic approach is able to describe correctly the minimum in $F$, it should also be able to describe the maximum in $G$. On the other hand, we have shown that clusters in stable/metastable equilibrium obtained in the NVT ensemble correspond with critical clusters in unstable equilibrium obtained in the NpT ensemble. As two faces of the same coin, this equivalence implies that one can infer the same information (e. g. the cluster radius, $\gamma$, or the nucleation rate) from either ensemble, as shown in the remainder of the paper. For this reason we will apply the Seeding method, previously used in the NpT ensemble (NpT-Seeding)[49,62,67], to the clusters equilibrated here in the NVT ensemble (NVT-Seeding). As discussed later on, the Seeding method uses information of the simulated clusters alongside CNT to provide estimates of $\gamma$ and the nucleation rate.

## V. ESTIMATING $\gamma$ FOR THE CLUSTERS

All the clusters obtained in the previous sections are in stable/metastable equilibrium in the NVT ensemble. According to the thermodynamic description presented in the book of Rowlinson and Widom[34] when the system reaches the equilibrium one obtains:

$$F = N\mu - p_{sol}V_{sol} - p_{liq}(V - V_{sol}) + \gamma A_{sol} \qquad (2)$$

where $p_{sol}$ and $p_{liq}$ are the pressures of a bulk solid and liquid respectively with chemical potential $\mu$. Notice that the chemical potential , and the temperature are homogeneous properties (the molecules can diffuse) whereas the density and pressure are inhomogeneous[76].

The way to proceed to evaluate $\gamma$ is as follows.

- The value of $F$ is computed

- A dividing surface of radius $R$ is chosen so that $V_{sol} = 4/3\pi R^3$ and $A_{sol} = 4\pi R^2$

- The value of $\gamma$ is obtained for the chosen dividing surface using Eq.2

Therefore, the value of $\gamma$ is not unique as it depends on the value of the chosen dividing surface. Two important surfaces are $R_e$ (for which the number of excess particles is zero) yielding $\gamma_e$; and $R_s$, giving rise to $\gamma_s$, which is the value for which $\gamma$ is minimum and for which the Laplace equation is satisfied.

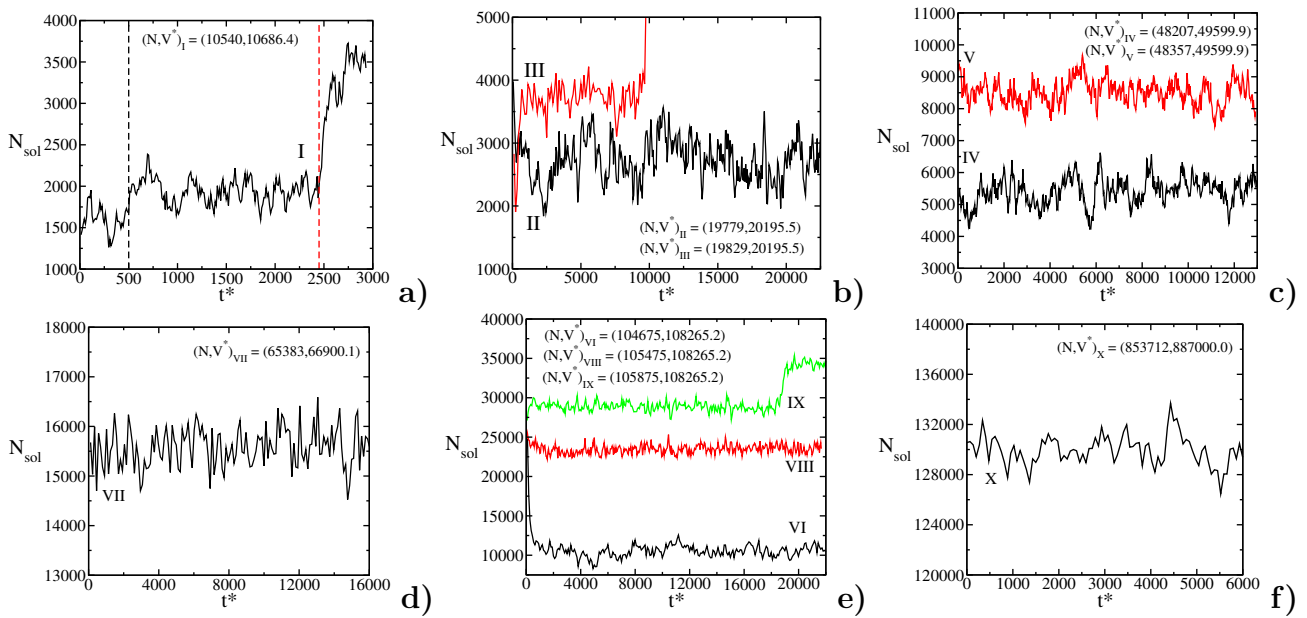

FIG. 3. Number of particles of the largest solid cluster as a function of time in reduced units. The clusters of each panel share the same volume. Hence, the difference in cluster sizes comes from the total number of particles, in other words, the net density. Details of the runs (volume, total number of particles, number of particles in the solid cluster, and pressure of the system while the spherical cluster is stable) are given in Table I. In panel a), the dashed line in the left indicates the starting point of the equilibrated coexistence of the spherical cluster within the fluid while the dashed line in the right indicates the transition to the cylindrical state.

As the reader may have noticed, the only way to determine $\gamma$ is to determine the free energy of the system. This has been done only in a couple of cases for the liquid-vapor interface by Binder and coworkers and can be regarded as a tour de force[2,5,21,23,27].

In the past, we have used an approximate approach denoted as Seeding[61] to circumvent the computation of free energies[49,62]. With NpT simulations of critical clusters alongside a "judicious" order parameter to label the particles as liquid and solid alongside CNT we obtained reasonable estimates of $\gamma$ and the nucleation rate[49,62]. By judicious we mean that the chosen order parameter is able to predict the free energy barrier and the nucleation rate obtained from rigorous techniques (for instance US[44,45] or metadynamics[77]). For HS, the combination of the order parameter of Lechner-Dellago $\bar{q}_6$[74] and the mislabeling criterion[61] meets this requirement[62]. According to CNT, the free energy barrier $\Delta G$ and the surface free energy $\gamma$ for the pressure $\langle p^* \rangle$ can be estimated by means of the following expressions:

$$\Delta G = \langle N_{sol} \rangle \frac{\Delta \mu}{2}, \qquad (3)$$

$$\gamma = \left( \frac{3 \langle N_{sol} \rangle \rho_{sol}^2 |\Delta \mu|^3}{32 \pi} \right)^{1/3}, \qquad (4)$$

where $\rho_{sol}$ is the density of the solid phase and $\Delta \mu \equiv \mu_{liq} - \mu_{sol}$ is the difference between the chemical potentials of the bulk liquid $\mu_{liq}$ and that of the bulk solid $\mu_{sol}$ when both are at the same pressure (this difference is not zero, as the equilibrium in the NVT ensemble arises from the higher pressure of the solid phase due to the presence of the curved interface). The variables $\rho_{liq}$ and $\rho_{sol}$ are obtained from the equations of state that are computed from simulations of the bulk phases along the isotherm of interest while $\Delta \mu$ is computed via thermodynamic integration $\Delta \mu = \int_{p_{coex}}^{p} [1/\rho_{sol}(p') - 1/\rho_{liq}(p')] dp'$ starting from the coexistence point where the chemical potential of both phases are equal[78]. The values of $\gamma$ thus obtained in this work from solid clusters equilibrated in the NVT ensemble (NVT-Seeding) are compared to those obtained in Ref.[49,62] (NpT-Seeding) in Fig. 8. As expected from Fig. 6, that shows the equivalence of clusters in both ensembles, $\gamma$ is the same in both cases.

The clusters obtained in this work in the NVT ensemble are at stable/metastable equilibrium. Therefore the value of $\gamma_s$ for a certain value of $R_s$ it is the value of the interfacial free energy obtained for this radius at the equilibrium pressure $\langle p^* \rangle$. Although one usually speaks on the variation of $\gamma_s$ with $R_s$ one should rather speak on the variation of $\gamma_s$ with the pair $R_s$ and $\langle p^* \rangle$ because it is not possible to change $R_s$ and $\langle p^* \rangle$ independently for a system that is at equilibrium.

## VI. VARIATION OF $\gamma$ WITH CURVATURE AND THE TOLMAN LENGTH

It should be emphasized that the values of $R$ and $\gamma$ that are used in CNT are those of the surface of tension, i.e. $R_s$ and $\gamma_s$[5,22,79]. In fact, it is simple to show that if one assumes that the chemical potential of the bulk liquid ($\mu_{liq}(p_{liq})$) is identical to that of the stable/critical solid cluster ($\mu_{sol}(p_{sol})$) (which makes sense after the results presented in the previous sections) and uses the Laplace equation (which is restricted to the surface of tension) to estimate the difference of pressures, one obtains the main equations of CNT (after assuming that the density of the solid does not change much with pressure). This indicates that $R_s$ and $\gamma_s$ are indeed the ones obtained when applying CNT.[5,22,79] Thus, values labeled as $R_c$ and $\gamma_{CNT}$ in our previous work should be identified with $R_s$ and $\gamma_s$.

It is obvious from the results of Fig. 8 that the value of $\gamma_s$ is not constant (showing the failure of the capillarity approximation). Recently, we used the following expression to describe the variation of $\gamma_s$ with the cluster radius[49]:

$$\gamma_s = \gamma_{0,T}\left(1 - 2\frac{\delta_T}{R_s}\right),\tag{5}$$

where $\delta_T$ is a fitting parameter and $\gamma_{0,T}$ is the inter-facial free energy at coexistence for a flat interface. We showed that this expression correctly describes the $\gamma$ variation for critical hard sphere clusters. The blue line in Fig. 8 is the fit obtained in Ref.[49]. It describes well the data coming from either ensemble, which further demonstrates the equivalence between clusters equilibrated in the NVT ensemble and critical clusters obtained in the NpT ensemble. The parameters are $\gamma_{0,T} = 0.576kT/\sigma^2$, the interfacial free energy at coexistence ($p^* = 11.648$) averaged over several planes[80,81], and $\delta_T = -0.41\sigma$. It is worth noting that, for HS in contact with a smooth spherical hard wall, a similar value of $\delta_T$ was reported from a theoretical study using Density Functional Theory[18] (although the value of $\gamma_{0,T}$ was found to be different indicating that there are differences in the value of $\gamma_{0,T}$ between a hard structureless spherical wall and a solid cluster of ordered hard spheres).

What is the physical meaning of the fitting parameter $\delta_T$? Since this parameter is a distance we can compare it with the Tolman length, $\delta_{Tolman}$[1,50]:

$$\delta_{Tolman} \equiv \lim_{\frac{1}{R_s}\to 0} \delta\tag{6}$$

where $\delta$ is the difference between the equimolar and the surface of tension radii:

$$\delta = R_e - R_s\tag{7}$$

The radius $R_e$ of the (spherical) Gibbs dividing surface is obtained simply from the equation

$$N = \rho_{sol}[(4/3)\pi R_e^3] + \rho_{liq}[V - (4/3)\pi R_e^3].\tag{8}$$

The previous expression only requires the knowledge of the bulk densities of the solid and fluid phases ($\rho_{sol}$ and $\rho_{liq}$ respectively) . On the other hand, $R_s$ can be calculated from:

$$R_s = [3\langle N_{sol}\rangle/(4\pi\rho_{sol})]^{1/3}.\tag{9}$$

The values of $\delta$ are reported in Table II. We found that $\delta$ is negative and its value changes with the radius of the solid cluster (i.e. with the equilibrium pressure). An analogous change of $\delta$ has been observed by Binder and coworkers in studies on the vapor-liquid interface[2,5,21,23,27]. In Fig.9 we have fitted the values of $\delta$ as a function of $1/R_s$ obtaining the value $\delta_{Tolman} = -0.41\sigma$ when $1/R_s$ goes to zero (i.e. for planar interface). The obtained value coincides with $\delta_T$ obtained from 5. Therefore, $\delta_T$ is an estimate of the Tolman length.

## VII. APPLICATION OF EQUILIBRIUM CLUSTERS TO STUDY NUCLEATION

We have developed in the last years an approximate route denoted as Seeding to determine nucleation rates, $J$. By performing NpT runs, the size of the solid critical

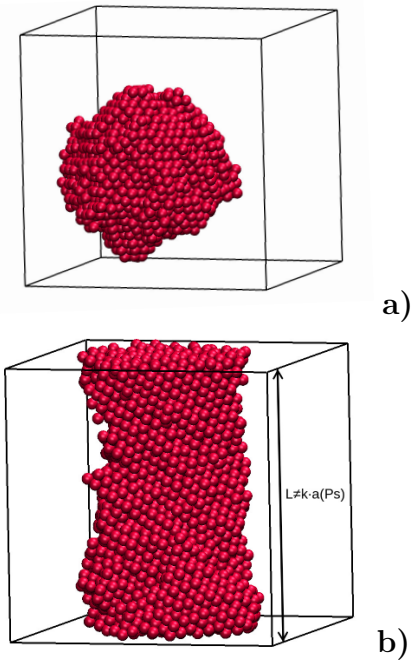

FIG. 4. Snapshots of: a) metastable sphere and b) stable cylinder within a fluid. Only crystalline particles are shown. In b), the vertical arrowed line indicates that, in general, the size of box does not necessarily meet $k$ times ($k$ being an integer) the length, $a$, of the unit cell of the FCC hard sphere crystal at the pressure $\langle p^*\rangle$. Thus, the solid may have some stress.

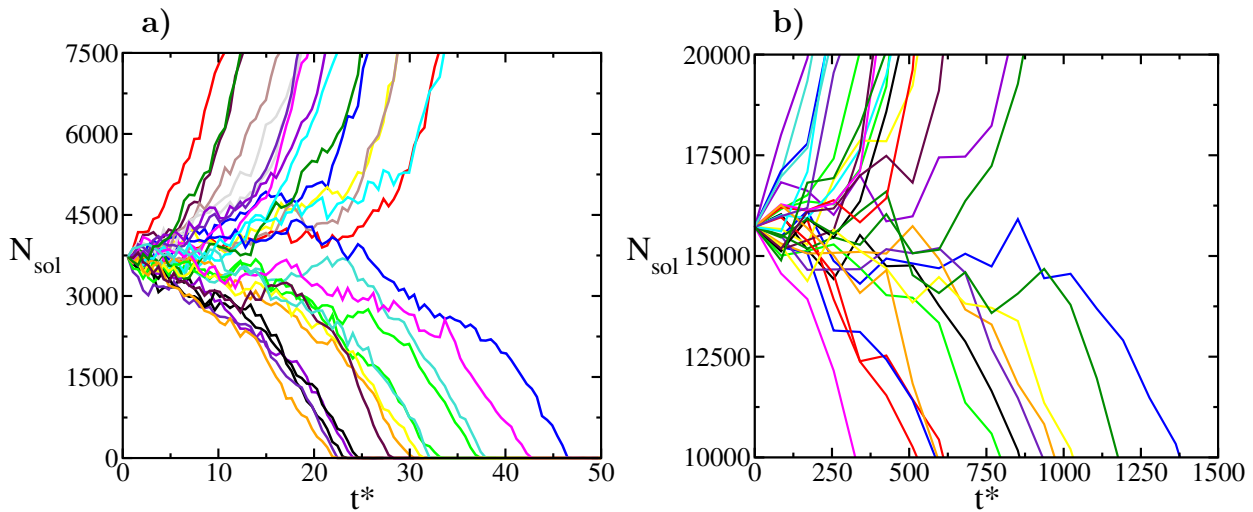

FIG. 5. Number of particles in the largest cluster within the system in an $NpT$ simulation at $p^*$ equal to the average pressure during the life time of the stable sphere in the $NVT$ run. The total number of runs is 30 in both cases. It can be seen how in $\sim 50\%$ of the trajectories the cluster either grows or disappears. The clusters correspond to cases III (left panel) and VII (right panel) of Table I.

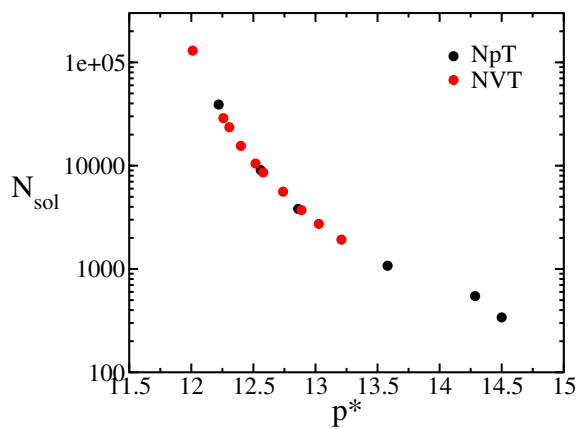

FIG. 6. Cluster sizes against pressure. Notice that although the stable clusters from NVT simulations were tested to be critical via NpT simulations, data labeled as NpT in this figure correspond to previous work of Seeding[49,62].

cluster $N_{sol}$ (at a certain $T$ and $p$) is determined, and $J$ is estimated from the expressions of CNT:

$$J = \rho_{liq}\sqrt{\frac{|\Delta\mu|}{6\pi kT N_{sol}}} f^+ exp[(-\Delta\mu N_{sol})/(2kT)], \quad (10)$$

where $f^+$ is the attachment rate which will be approximated as

$$f^+ = 24 D_{liq}(N_{sol})^{(2/3)}/\lambda^2, \quad (11)$$

where $\lambda$ is the attachment length which for HS can be approximated[62] as $\lambda \simeq (\sigma/4)$ and $D_{liq}$ the diffusion coefficient of the fluid at the pressure $p$. In previous work, we have shown that this set of equations (with the input from simulations) provides an excellent description of the values of $J$ (including those for HS). We could denote this approach as NpT-seeding as a number of runs are performed at $N$, $p$, and $T$ constant.

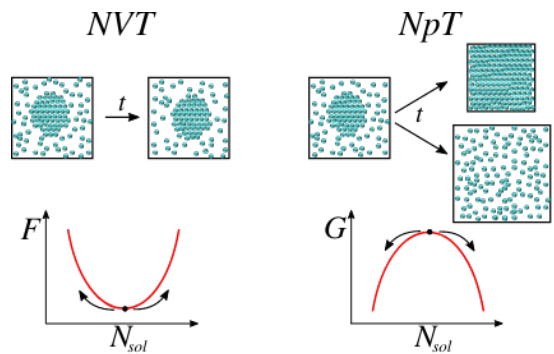

FIG. 7. Sketch of the free energy profile for the case of a stable solid cluster composed by $N_{sol}$ particles in the $NVT$ ensemble (left). The system is in a free energy minimum of the Helmholtz free energy $F$ and does not change with time (except for thermal fluctuations). By switching to the $NpT$ ensemble, the system ends up in a maximum in the Gibbs free energy $G$ (right) and evolves either towards to the solid phase or towards the fluid phase with the same probability. In both cases, the value of $\gamma$ is the same. Equilibrium (left) and nucleation (right) can be regarded as the two faces of the same coin.

However, the results of Fig.5 indicate that there is a new way of doing Seeding. Instead of inserting a solid cluster in an equilibrated fluid and performing a number of NpT runs to determine at which pressure the cluster is critical, one can equilibrate the solid cluster in the NVT ensemble. In this way, the size of the solid cluster at

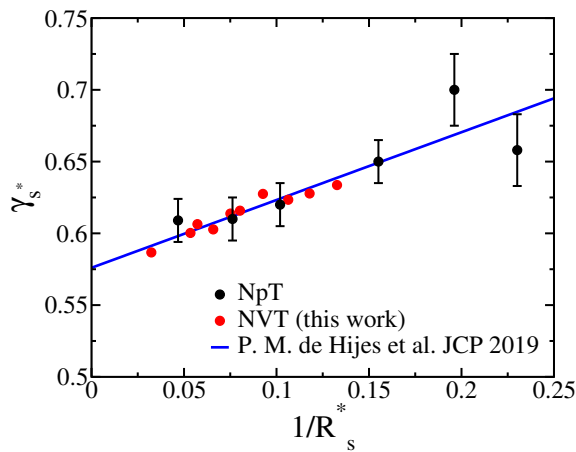

FIG. 8. Interfacial free energy against the inverse of the radius of the cluster from both equilibrium results of this work (labeled as $NVT$) and from nucleation studies of our previous work[49,62] (labeled as $NpT$). Also shown: linear expression proposed in previous work[49] in solid blue line.

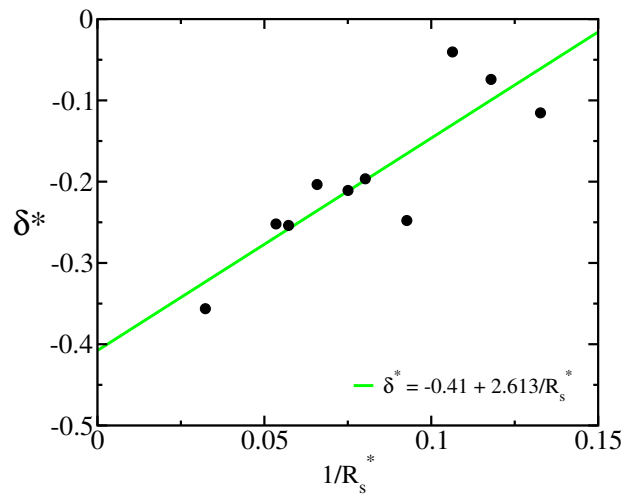

FIG. 9. $\delta^*$ as function of $1/R_s^*$ for the different stable clusters of the $NVT$ runs. Green dashed curve is a linear fit of the data.

| Label | $R_s^*$ | $R_e^*$ | $\delta^* \equiv (R_e^* - R_s^*)$ |
|-------|---------|---------|-----------------------------------|
| I     | 7.535   | 7.419   | -0.1154 |
| II    | 8.481   | 8.407   | -0.0742 |
| III   | 9.402   | 9.362   | -0.0404 |
| IV    | 10.791  | 10.543  | -0.2478 |
| V     | 12.461  | 12.265  | -0.1965 |
| VI    | 13.322  | 13.111  | -0.2108 |
| VII   | 15.200  | 14.996  | -0.2035 |
| VIII  | 17.467  | 17.213  | -0.2538 |
| IX    | 18.700  | 18.448  | -0.2521 |
| X     | 30.927  | 30.571  | -0.3563 |

TABLE II. Radius associated to the dividing surfaces from our results from equilibrium clusters. $R_s$ is computed with Eq. (9) and corresponds to the radius of the sphere containing a number of particles corresponding to the equilibrium one, as reported in Tab. I. $R_e$ corresponds to the sphere's radius of the equimolar dividing surface (Gibbs dividing surface). $\delta$ is, by definition, the difference between the previous radii.

the pressure $\langle p^* \rangle$ is obtained from a single equilibrium run. After this is done, one can use the machinery of Seeding (or more precisely Eq.10) to estimate $J$. We shall denote this approach as NVT-Seeding. We shall now estimate nucleation rates, simply using the results for the equilibrium solid clusters presented in Table I (plus performing additional simulations to estimate $\Delta\mu$ and $D_{liq}$). All the results required to determine $J$ from the equilibrium solid clusters of this work are presented in Table III

Values of the nucleation rate $J$ computed in this work are presented in the last column of Table III, whereas in Fig. 10 we compare them with our previous work[49,62] as well as other numerical[44,46] and experimental[82-85] independent estimations. Given that $J$ goes to zero when the pressure tends to its coexistence value, in Fig. 10, we show only the highest nucleation rates. The results obtained for $J$ from the equilibrium clusters of this work, agree quite well with previous results obtained from simulation techniques[44,46,47,62,86]. However the results of this work clearly contradict those found in experiments, providing further evidence that the experimental values presented as homogeneous nucleation rates are probably affected by heterogeneous nucleation events, as recently suggested in Ref.[87]. Previously mentioned nucleation studies of HS sampled the region of high pressures, typically above $p^* > 15$ (i.e. $\phi = (\pi/6)\rho^* > 0.52$). The results of this work expand the study to lower pressures (i.e. between $p^* = 12$ and $p^* = 13.2$ (i.e. $0.5 < \phi < 0.515$), closer to the coexistence pressure.

The NVT-seeding approach does not only work for nucleation of solid HS. We have also shown recently that this approach is also working for an entirely different problem (the cavitation of a bubble in a Lennard-Jones fluid at negative pressures)[30]. This NVT-seeding approach allows to study easily nucleation along isotherms (NpT-seeding can be implemented easily both along isobars and along isotherms). However, it has two drawbacks. The first one is that it cannot be applied to small solid clusters as it is impossible to have them in a stable configuration in the NVT ensemble. The second is that there may be finite size effects, as the ratio $\langle N_{sol}\rangle/N$ cannot be changed at will. For instance for a solid cluster of HS at equilibrium in the NVT ensemble with $\langle N_{sol}\rangle = 3200$ particles we found $\langle N_{sol}\rangle/N = 0.16$. In the NpT-seeding approach, this ratio can be made arbitrarily low (we typically set it to $\langle N_{sol}\rangle/N < 0.05$ in our previous work). We found that the pressure at which the cluster was critical in the NVT system was $p^* \sim 12.95$ whereas it was found to be $p^* \sim 13.05$ in the NpT ensemble when $\langle N_{sol}\rangle/N$ was small. The finite size effects on HS is not dramatic (less than one per cent for the pressure at which the cluster is critical) but one should be aware of their presence.

| Label | $\rho_{liq}^*$ | $\rho_{sol}^*$ | $|\Delta\mu|/kT$ | $\gamma^*$ | $\Delta G/kT$ | $D_{liq}/(\sigma^2/\tau)$ | $f^+/(6D_{liq}/\sigma^2)$ | $\log_{10}[J/(6D_{liq}/\sigma^5)]$ |
|---|---|---|---|---|---|---|---|---|
| I | 0.970 | 1.074 | 0.1566 | 0.634 | 150.7 | 0.0183 | 9904 | -63 |
| II | 0.967 | 1.071 | 0.1383 | 0.628 | 189.1 | 0.0190 | 12520 | -79 |
| III | 0.964 | 1.068 | 0.1242 | 0.623 | 230.9 | 0.0195 | 15360 | -97 |
| IV | 0.962 | 1.065 | 0.1092 | 0.628 | 306.0 | 0.0201 | 20192 | -130 |
| V | 0.959 | 1.061 | 0.0931 | 0.617 | 400.5 | 0.0207 | 26869 | -171 |
| VI | 0.958 | 1.060 | 0.0869 | 0.614 | 456.3 | 0.0209 | 30684 | -195 |
| VII | 0.956 | 1.057 | 0.0750 | 0.603 | 583.3 | 0.0213 | 39879 | -250 |
| VIII | 0.954 | 1.055 | 0.0658 | 0.606 | 775.0 | 0.0217 | 52594 | -333 |
| IX | 0.953 | 1.054 | 0.0609 | 0.600 | 879.3 | 0.0219 | 60242 | -378 |
| X | 0.949 | 1.049 | 0.0362 | 0.587 | 2350.6 | 0.0228 | 164176 | -1017 |

TABLE III. Results from NVT-seeding calculations as obtained from the simulations of this work..

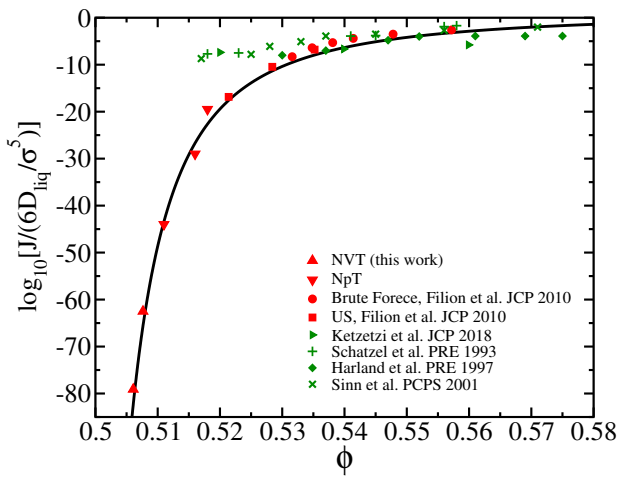

FIG. 10. Nucleation rate against volume fraction (defined as $\phi \equiv \rho^*\pi/6$) from different sources. Experimental and numerical results are shown in green and red, respectively. The black line is a fit using CNT equations accounting for the interfacial free energy variation with the radius of the cluster as proposed in Ref.[49]. As stated in the caption of Fig. 8 points labeled as NpT in the legend come from previous work[49,62].

## VIII. CONCLUSIONS

In this work, we extend the simulations pioneered by Binder and coworkers[2,5,19–27] and recently used by Richard et al. also for HS to study a curved interface[14]. In particular, we have considered the possibility of having a stable spherical solid cluster in equilibrium with a fluid, for the hard spheres system. We were able to obtain equilibrium for up to ten different clusters with sizes ranging from 1900 to 130000 particles.

After the equilibrium configuration was found, we launched NpT runs and found that the clusters were critical at the average pressure found in the NVT run. Accordingly, all properties that can be inferred from critical clusters in unstable equilibrium with the fluid in the NpT ensemble coincide with those obtained from clusters in stable/metastable equilibrium in the NVT ensemble. We show this equivalence for the cluster radius as well as for $\gamma$ and the nucleation rate obtained from a Seeding analysis (CNT fed by microscopic parameters of the clusters measured in the simulations). Therefore, whereas the system is in a minimum of $F$ in the NVT ensemble, the fact that the solid cluster is critical indicates that the system is in a maximum of $G$ in the NpT ensemble. This is in agreement with a recent NVT-Seeding study of bubble cavitation[30].

In addition, we study the variation of $\gamma$ with $R_s$, the relevant dividing surface in CNT. Recently, we showed by means of simulations of critical clusters in the NpT ensemble that such variation is well described by a linear fit of $\gamma$ versus $1/R_s$ and obtained a characteristic length $\delta_T$ as a fitting parameter[49]. In this paper we show that the fit obtained in Ref.[49] works well for clusters equilibrated in the NVT ensemble as well. Moreover, we obtain the Tolman length as the difference between $R_e$, the Gibbs dividing surface, and $R_s$ in the limit of very large clusters. We obtain $R_e - R_s$ for the clusters equilibrated in the NVT ensemble and extrapolate the difference to infinite radius. With this procedure we estimate the Tolman length, $\delta_{Tolman}$. We find that $\delta_{Tolman}$ coincides with the $\delta_T$ parameter obtained from the fit of $\gamma_s$ versus $(1/R_s)$ above mentioned.

We hope this work will encourage further research on the fascinating (but arguably difficult) issue of the change of the interfacial free energy between two phases separated by a curved interface.

## IX. ACKNOWLEDGMENTS

This work was funded by Grant FIS2016-78117-P of the MEC and by project UCM-GR17-910570 from UCM. P. M. de H. acknowledges financial support from the FPI grant No. BES-2017-080074. J. R. E. acknowledges funding and support from the Oppenheimer Research Fellowship and from the Emmanuel College Research Fellowship. V.B. acknowledges the support from the European Commission through the Marie Skłodowska-Curie Fellow-ship No. 748170 ProFrost. C.V would like to thank Dr. Mario Alvarez and Dr. Aurora Rodriguez for their help in the last two years. Without them this work would have not been possible. We would like to dedicate this paper to the memory of Prof. John Rowlinson who passed away in 2018.